\documentclass[a4paper]{jpconf}
\usepackage{graphicx}
\begin{document}
\title{Superfluid hydrodynamics in fractal dimension space}

\author{D.A. Tayurskii and Yu.V. Lysogorskiy}

\address{Institute of Physics, Kazan University, 18 Kremlevskaya st., Kazan, 420018, Russia.}

\ead{dtayursk@gmail.com}

\begin{abstract}
The complex behavior of liquid ${}^4$He and liquid ${}^3$He in nanoporous media is determined by influence of randomly distributed geometrical confinement as well as by significant contribution from the atoms near walls. In the present paper fractional Schrodinger equation has been used for deriving two-fluid hydrodynamical equations for describing the motion of superfluid helium in the fractal dimension space. Nonlinear equations for oscillations of pressure and temperature are obtained and coupling of pressure and temperature oscillations is observed. Moreover coupling should disappear at very low temperatures which provide an experimental test for this theory.
\end{abstract}

\section{Introduction}

Liquid helium  ${}^4$He  and ${}^3$He at very low temperature belong to the class of
quantum fluids with strong correlations between atoms. The first one represents a Bose-system and shows superfluid transition at $\lambda$-point $T_\lambda=2.17$ K (below this temperature it is called He-II). In recent years it has been recognized that the quantum fluids in the confined geometry at nanoscale length can be considered as a new state of quantum matter due to close values of characteristic lengths for these quantum liquids and the size of geometrical confinement as well as due to significant contribution to the physical properties from the atoms near walls.

The question of influence of geometrical factors (confinement dimension, dimension of nanopores space, etc) has been emerged\cite{Tayurskii1,Tayurskii2,Wong,Matsumoto,Vasquez,Azuah}. So one has to
apply a new physics to describe such systems with taking into
account their complex nature. For example, last two years the
attempts to develop the fractionalized two-fluid hydrodynamics for
nanoporous media with fractal dimensions have been
made\cite{TayuLys1,TayuLys2}.

In the present paper fractional Schrodinger equation has been used for deriving two-fluid hydrodynamical equations for describing the motion of superfluid helium in the fractal dimension space (like aerogel) and nonlinear equations for oscillations of pressure and temperature are obtained and coupling of pressure and temperature oscillations is observed.

\section{Two-fluid hydrodynamical model}
The usual two-fluid hydrodynamic model ~\cite{Tilley},~\cite{Khalatnikov} of superfluid helium ${}^4$He is described by the system of four differential equations~(\ref{continuity}), (\ref{entropycontinuity}), (\ref{superfluideuler}), (\ref{normaleuler}) which represent continuity equation, entropy conservation law and Eiler equations for superfluid and normal components respectively
\begin{eqnarray}
\label{continuity} \frac{\partial\rho}{\partial t}&+&\nabla{\textbf
j}=0 \\
\label{entropycontinuity}
\frac{\partial\rho\sigma}{\partial t}&+&\nabla(\rho\sigma {\textbf
v}_n)=0\\
\label{superfluideuler}
\rho_s \frac{\partial{\textbf v}_s}{\partial
    t} &=&
    -\frac{\rho_s}{\rho}\nabla p+\rho_s\sigma\nabla T \\
\label{normaleuler}
\rho_n \frac{\partial{\textbf v}_n}{\partial
    t} &=&
    -\frac{\rho_n}{\rho}\nabla p-\rho_s\sigma\nabla T,
\end{eqnarray}
where $\rho$~-- He-II density, $\rho_s,\rho_n$~-- superfluid and
normal component density, ${\textbf v}_s,{\textbf v}_n$~--
superfluid and normal component velocity, ${\textbf
j}=\rho_s{\textbf v}_s + \rho_n{\textbf v}_n$ -- He-II flow
density. Interaction between the fluid components, forces of viscous friction and energy dissipation are neglected because of the small velocities of both components. This model provides two type of oscillations - oscillations of pressure (first sound, both components move in phase) and oscillations of temperature (second sound, components move in counter-phase).  Neglecting the anomalously small thermal expansion coefficient of He-II, oscillations of pressure and temperature in bulk sample of superfluid helium are independent. But in superfluid helium in aerogel the experimentally proven coupling of these two type of oscillations exists\cite{mckenna}.

\section{Fractional quantum mechanics}
To take into account complex nature of internal geometry of confinement (aerogel) it is reasonable to use the formalism of fractional quantum mechanics. In the framework of the so-called Feynman formalism for quantum mechanics the key concept is the concept of trajectory and a particle can move along any possible trajectories. To move a particle from point A to point B one has to take into account the contributions from all possible trajectories with the corresponding weight (complex factor). Possible trajectories resemble Brownian trajectory of a free particle and have a fractal dimension of $\alpha$ = 2.

De Broglie thermal wavelength for the helium atom inside the aerogel at temperatures of about 1 K is about 10 \AA, which is in proper relation with the characteristic length scale of the fractal structures formed aerogel. Thus, some quantum-mechanical trajectory of the helium atom will be forbidden due to the influence of the structure of aerogel. The realized trajectory will resemble the motion of a Brownian particle in a porous medium, where the mean square displacement depends on time as $<x^2>\propto t^{\alpha} $. This phenomenon is called subdiffusion. To describe this type of diffusion an equation with fractional Riesz derivative is used. Probability density function for such case is written in terms of Levy function, which is a generalization of the Gaussian distribution\cite{PhysOfFractalOperator2003,Laskin}.

It has been proposed to generalize the Feynman's path integrals to an arbitrary fractal dimension of trajectories $\alpha$ \cite{Laskin}. From this type of path integrals one can obtain fractional Schr\"{o}dinger equation and fractional Hamiltonian in the following form
\begin{eqnarray}
\label{FSE}
 i\hbar\frac{\partial \psi(\textbf{r},t)}{\partial t}&=&\hat{H}_\alpha \psi(\textbf{r},t)\\
\label{FracHamiltonian}
\hat{H}_\alpha&=&D_\alpha \left(\hbar \nabla\right)^{\alpha}+V\left(\textbf{r},t\right) = D_\alpha \left|\hat{\mathbf{p}}\right|^\alpha+V\left(\textbf{r},t\right),
\end{eqnarray}
where $\left(\hbar \nabla\right)^{\alpha}$ is Riesz fractional operator\cite{Laskin}. Thus fractional Hamiltonian~(\ref{FracHamiltonian}) is the Hermitian operator and provide probability conservation law. Also it obeys parity conservation law, so one divide particles at the fermoins and bosons, as in usual case. This type of Hamiltonian~(\ref{FracHamiltonian}) already has been used to describe the specific heat of noncrystalline solids (glasses) associated with the unusual structure of these materials\cite{Lenzi}.

\section{Galilean noninvariance}
On other hand, the non-quadratic form of the Hamiltonian ~(\ref{FracHamiltonian}) with respect to momentum operator results in breaking some other symmetries in dynamic equations, for example, the Galilean invariance. In usual case, transition from one inertial system to another will conserve general view of dynamical equations. If we consider Schr\"{o}dinger equation in two inertial reference frame $K$ (nonprimed variables) and $K^\prime$ (primed variables), which move relative to each other with velocity $V$, then by special substitution\cite{Greenberger} one can achieve the same form of dynamical equation. On other hand, by making similar substitution in the form $\Psi(x,t)=\varphi(x^\prime,t) e^{i \frac{m V x^\prime}{\hbar}+i \frac{D_\alpha \left|p\right|^\alpha t}{\hbar}}$, for fractional Schr\"{o}dinger equation it is impossible to get the similar form of equation as in reference frame $K$. The only possibility to resolve it to set $V=0$, i.e. do not change the intertial reference frame. As a consequence there is some special reference frame where the fractional Schr\"{o}dinger equation has form~(\ref{FSE}). One can identify this reference frame with some reference frame where, for example, nanoporous media is in a rest.

\section{Fractional two-fluid hydrodynamical model}
Within the Heisenberg representation the dynamical equations for operators $\hat{\mathbf{r}}$ and $\hat{\mathbf{v}} = \frac{d\hat{\mathbf{r}}}{dt}$ are written as
\begin{eqnarray}
\label{GNFracVelocity}
\widehat{\mathbf{v}} &=& \alpha D_\alpha \left|\widehat{\mathbf{p}}\right|^{\alpha-2}\widehat{\mathbf{p}},\\
\label{GNFracNewtonLaw}
\frac{d\widehat{\mathbf{v}}}{dt} &=& -\frac{i}{\hbar} \alpha D_\alpha \sum_{l=1}^{\infty} \frac{(-i \hbar)^l}{l!} \nabla^l V(\widehat{\mathbf{r}}) \frac{(\alpha-1)!}{(\alpha-l-1)!} \widehat{\mathbf{p}}^{\alpha-1-l},
\end{eqnarray}
where $V$ represents potential energy. Suppose that velocity is more general quantity than momentum~\cite{Dong2008}, one can rewrite~(\ref{GNFracNewtonLaw}) in terms of velocity and leave only the first spatial derivation of potential $V$.   From fractional Schrodinger equation the mass conservation law follows in the form
 \begin{equation}
\label{FHConservationLaw}
\frac{\partial \rho(\mathbf{r},t)}{\partial t}+\mathbf{\nabla}\mathbf{J}(\mathbf{r},t)+K(\mathbf{r},t) = 0,
\end{equation}
where $\rho = \Psi^* \Psi$ is the probability density, $\mathbf{J} = \frac{1}{\alpha}\left(\Psi^* \hat\mathbf{v}\Psi + \Psi \hat\mathbf{v}\Psi^*\right)$ is the density flow\cite{Laskin} and $K = \frac{i}{\alpha \hbar}\left( \hat{\mathbf{p}} \Psi \hat{\mathbf{v}} \Psi^* - \hat{\mathbf{p}} \Psi^* \hat{\mathbf{v}} \Psi  \right)$
 is a new term, which represents additional sources in fractal space. If we assume that density of superfluid helium is almost homogeneous, i.e. we have only set of plane waves with close values of $\mathbf{p}$ in wave function and helium atoms is strongly delocalized, then one can suppose $K \approx 0$.
In that case in fractional two-fluid hydrodynamical model one can keep the continuity equations of mass and entropy in the form~(\ref{continuity}) and~(\ref{entropycontinuity}).

  Instead of parametr $D_\alpha$ we introduce a new one with dimension of velocity as $\alpha D_\alpha = v_0^{2-\alpha} m^{1-\alpha}.$ Then dynamic equation~(\ref{GNFracNewtonLaw}) for the superfluid component with taking into account thermodynamic relations~\cite{Tilley} can be rewritten as
  \begin{equation}
  \label{FHFracSuperfluidEilerFinal}
\rho_s \frac{d \mathbf{v}_s}{dt} = (\alpha-1)\left|\frac{\mathbf{v}_s}{v_0}\right|^\frac{\alpha-2}{\alpha-1}\left(-\frac{\rho_s}{\rho}\mathbf{\nabla} p + \rho_s \sigma \mathbf{\nabla} T\right).
  \end{equation}
  Also we can suppose the similar form for dynamic equation to which the density flow obeys and write
\begin{equation}
  \label{FHFracTotalJEquation}
\frac{d \mathbf{j}}{dt} = -(\alpha-1) v_0^\frac{2-\alpha}{\alpha-1} \left|\frac{\mathbf{j}}{\rho}\right|^\frac{\alpha-2}{\alpha-1} \mathbf{\nabla} p.
\end{equation}

Thus equations~(\ref{continuity}), (\ref{entropycontinuity}), (\ref{FHFracSuperfluidEilerFinal}) and~(\ref{FHFracTotalJEquation}) form \emph{fractional two-fluid hydrodynamical model} of superfluid helium in nanoporous space. This set of  equations results in two oscillation equations for pressure and temperature. By using  "weak fractality" approximation $\alpha-2 \ll 1$ in low temperature region $T<0.5$K when $\rho_n/\rho_s \ll 1$ and $v_n/v_s \ll 1$ one can rewrite these equations up to $O(\alpha-2)$ as
\begin{eqnarray}
\label{FHPressureOsc}
\frac{\partial^2 p}{\partial t^2} &=& u_1^2\left(1+(\alpha-2)(1+\ln \left|\frac{v}{v_0}\right|)\right) \nabla^2 p+u_1^2 (\alpha-2) \nabla \ln \left|\frac{v}{v_0}\right| \nabla p,\\
\label{FHTemperatureOsc}
\frac{\partial^2 T}{\partial t^2} &=& (\alpha-1) u_2^2 \left(1+(\alpha-2) \ln\left|\frac{v_s}{v_0}\right|\right) \nabla^2 T +(\alpha-2)u_2^2 \frac{\rho_n}{\rho \sigma \rho_s} \left(\frac{v_n}{v_s}-1\right)\nabla^2 p,
\end{eqnarray}
where $u_1$ and $u_2$ is first and second sound velocity for bulk superfluid helium. We will look for the solutions in the form of plane waves $p = p_0 + p^\prime e^{i\left(u k t - k r\right)}$ and $T = T_0 + T^\prime e^{i\left(u k t - kr\right)}$, where $u$ is sound velocity, and suppose that helium velocity has dependence $v=v^0 \exp(i(u k t - k r))$. As a result one can obtain two type of oscillations: temperature oscillations with sound speed $u \approx u_2 \left(1+(\alpha-2)(1+\frac{1}{2}\ln \left|\frac{v_s^0}{v_0}\right|)\right)$  and pressure-temperature oscillations with sound speed $u \approx u_1 \left(1+\frac{\alpha-2}{2}(1+\ln \left|\frac{v^0}{v_0}\right|)\right)$. The latter one has the relation between pressure and temperature amplitudes $T^\prime = \beta p^\prime$, where coupling coefficient has form $\beta = ((\alpha-2) \frac{u_2^2 \rho_n}{\rho \sigma \rho_s} \left(\frac{v_n^0}{v_s^0}-1\right))/(u_2^2 - u_1^2)$. At low temperatures, when $u_1^2 \approx \textrm{const}$, $u_2^2 \approx \textrm{const}$ and $\sigma \propto T^n$, this coupling coefficient should have linear dependence on temperature $\beta \propto T$ which provide us with possible experimental test for proposed model.

\section{Conclusion}

It was proposed that for the microscopical description of superfluid in nanoporous media with complex fractal structure one can use fractional Schr\"{o}dinger equation. But it is necessary to keep in mind that the fractal geometry of nanoporous media leads to the Galilean
noninvariance of this equation and as a consequence one needs to choose the special frame of reference where, for example, nanoporous media is in a rest. One can interpret dynamic equations of fractional quantum operator in Heisnberg representations as a classical dynamic equations in fractal media and generalize them to obtain fractional hydrodynamic set of equations. From this two-fluid fractional hydrodynamic model one can obtain equations of pressure-temperature and pure temperature oscillation. It was shown that the pressure-temperature coupling constant has linear dependence on temperature at low temperature region which provide us with possibility of experimental proof of given model.

\ack
This work is supported in part by the Russian Fund for the Fundamental Research (09-
02-01253)

\end{document}